\documentclass[twocolumn,showpacs,preprintnumbers,amsmath,amssymb]{revtex4}

\usepackage{graphicx}
\usepackage{epsfig}
\begin{document}
\newcommand{\be}{\begin{equation}}
\newcommand{\ee}{\end{equation}}
\newcommand{\bea}{\begin{eqnarray}}
\newcommand{\eea}{\end{eqnarray}}
\newcommand{\bc}{\begin{center}}
\newcommand{\ec}{\end{center}}

\preprint{FT-02-15}

\title{Dynamical transition between two mesons and a tetraquark.}

\author{I. A. Toledano Ju\'arez and G. Toledo S\'anchez}

\affiliation{ Instituto de F\'{\i}sica, Universidad Nacional Aut\'onoma de M\'exico. M\'exico D. F. C.P. 04510}

\date{\today}
            
\begin{abstract}
 We consider a system composed of two identical light quarks ($qq$) and two identical antiquarks ($\bar Q\bar Q$) that can be linked either as two mesons or as a tetraquark, incorporating quantum correlations between identical particles and an effective many-body potential between particles. We perform a 3-D Monte Carlo simulation of the system, considering the configurations allowed to form: i) Only two mesons,  ii) Only  tetraquark  and iii) two mesons and tetraquark . We characterize each case and determine whether it is energetically more favorable to form a tetraquark or two mesons, as a function of the interparticle separation distance which, for a fixed number of particles, can be identified as a particle density.
We determine how the two mesons, which dominate the low density regime, mixes with a tetraquark state as the density increases. Properties like the mean square radius and the two-particle correlation function are found to reflect such transition, and we provide a parameterization of the diquark correlation function in the isolated case. We track the dynamical flipping among configurations to determine the recombination probability, exhibiting the importance of the tetraquark state. We analize the four-body potential evolution and show that its linear behavior is preserved, although the slope can reflect the presence of a mixed state.
Results are shown for several light-quarks to heavy-antiquarks mass ratios whenever they are found to be relevant.

\end{abstract}

\maketitle

\section{Introduction} 
The formation of multi-quark systems can have important implications in the phenomena we observe in nature, from an enhanced spectroscopy to quark recombination effects. A tetraquark state, the simplest of this kind of systems, has been extensively studied from the theoretical and experimental point of view. Recently, experimental research has provided strong evidence on the formation of such state \cite{Belle,Bes,lhcb}. These findings signal a new era in hadron spectroscopy, where more refined modeling and a better understanding of the strong interaction, in the low energy regime, is required.
Since the early years of the quark model, theoretical studies have been performed to inquiry on the existence and stability of the tetraquark as an isolated object \cite{Jaffe,isgur83,isgur85,lenz,carlson,stancu94,Val,shilinzhu,T1,CFT,GHM,santopinto,reviewIJMPA}, and how its mixing with a meson state can help us to understand the observed spectroscopy of states like the $\sigma$ meson \cite{reviewIJMPA, mixing}.  
Less attention has been paid to the features of the tetraquark formation as two mesons are forced to approach each other as it could happen in a meson-meson collision or, on the opposite, when the four quarks are produced very close in space as in the $WW$ decay, which eventually freeze out to two mesons \cite{wmass,ww}. At which stage they turn into a tetraquark or mixed state? and how their properties reflect such modification? are certainly important questions.\\
In the present work we address these questions using an effective model (String-Flip Model)  to mimic the strong interaction among quarks. For definiteness we consider two identical light quarks $qq$ (denoted by $u$) and two identical antiquarks $\bar Q\bar Q$ (denoted by $d,s,c$ and $b$, depending on the mass ratio respect to the light  one). The seminal work by Lenz {\it et al.} \cite{lenz} laid down a procedure to describe such a system in this context and found its general properties, as isolated objects, using a harmonic potential interaction. Here, we perform a 3-D Monte Carlo simulation of the system including quantum correlations between particles to determine whether it is energetically more favorable to form a 4-body state (tetraquark) or two mesons, as a function of the interparticle separation which, for a fixed number of particles, can be identified as a particle density. We are interested in s-wave states where tensor and spin interaction effects are expected to be negligible in the gross features of the properties.
In section II, we set the basis to identify a quark-antiquark state interacting via a linear potential and elaborate on a variational approach. In section III, we describe the dynamical quark recombination model and generalize the variational two-body wavefunction to a four-body system. In section IV, we describe the energy evaluation from Monte Carlo simulation. In section V, we show the results on: A) The determination of the optimal variational parameter and system energy, B) The hadron radial distribution characterization, C) The two-particle correlation functions,  among the two mesons and among quarks. In particular we provide a parameterization of the diquark correlation function and the static structure factor,  D)  The dynamical recombination probabilities among mesons with/without tetraquark configurations, E) We characterize the four body potential, estimating its linear behavior and  its strength as an effective contact interaction. The conclusions are presented in section VI. 

\section{Quark-antiquark state with linear potential}
Let us start with the description of the exact solution and the variational approach of a meson state, composed of a quark and an antiquark of mass $m_1$ and $m_2$ respectively. The strong interaction between this pair can be represented by a flux tube \cite{isgur85}, which can be effectively described by a linear potential $V \left[ \vec{r}_1, \vec{r}_2 \right] = k \left\vert \vec{r}_1 - \vec{r}_2 \right\vert = k r$ where $k$ is an interaction constant and $r=|\vec{r}_1 -\vec{r}_2|$ is the relative distance between them. Lattice QCD studies have confirmed the linear behavior of the interaction among quarks at large distances \cite{qqlinear}.

\subsection{Exact solution}
 For a three-dimensional space, one can analytically solve the corresponding Schr\"odinger equation:
\begin{equation}
\left[ \frac{P_1^2}{2 m_1}+ \frac{P_2^2}{2 m_2} + k r \right] \Psi \left( \vec{r}_1,\vec{r}_2 \right) = E \ \Psi \left( \vec{r}_1,\vec{r}_2 \right). 
\end{equation}

\noindent Since the potential depends only on the relative distance, we can use the  center of mass ($\vec R$) and  relative distance ($\vec r$) vectors:
\begin{equation}
\vec{R} \equiv \frac{m_1 \vec{r}_1 + m_2 \vec{r}_2}{m_1 + m_2}; \quad \vec{r} \equiv \vec{r}_1 - \vec{r}_2,
\label{Coordr}
\end{equation}
and the wave function can be set as:
\begin{equation}
\Psi \left( \vec{R},\vec{r} \right) = \Phi(\vec{R}) \rho (\vec{r}).
\end{equation}
We can neglect the center of mass contribution and focus in the relative distance Schr\"odinger equation. 
Considering a spherical symmetry ($l$=0), the radial solution can be set as $U(r) = r \rho(r)$, such that the radial equation becomes:
\begin{equation}
-\frac{1}{2 \mu} \frac{d^2 U(r)}{dr^2} + k r U(r) = E U(r),
\label{Schr1}
\end{equation}
where $\mu \equiv {m_1 m_2}/({m_1 + m_2})$ is the reduced mass. 
Switching to  the variable 
\begin{equation} \label{cvar}
\xi \equiv \left(  r - \frac{E}{k} \right) \left( 2 \mu k \right)^{1/3};  \  \ \eta(\xi) =U(r)
\end{equation}

\noindent The  Schroedinger Eqn. (\ref{Schr1}) takes the following form:
\begin{equation} 
\frac{d^2}{d \xi^2} \eta(\xi) - \xi \eta(\xi)  = 0,
\label{Airy}
\end{equation}
whose general solution is a linear combination of the Airy functions
$\mbox{Ai}(\xi)$ and $\mbox{Bi} (\xi)$.  The second function is not considered as it has a bad behavior at large distances ($\xi \rightarrow \infty$, $\lim \mbox{Ai}(\xi) = 0$ and $\lim \mbox{Bi}(\xi) = \infty$). Thus, the solution is:
\begin{equation} \label{eq4}
\eta(\xi) = \mbox{Ai}(\xi)
\end{equation}
The boundary condition at the origin $U(0) = 0$ requires
\begin{equation}
 \eta \left(  - \frac{E}{k} \left( 2 \mu k \right)^{1/3} \right) = 0,
\label{Airy3}
\end{equation}
corresponding to the zeros of the Airy function $\xi_n$, which then define the eigen-energies $E_n$:
\begin{equation}
E_n = \left[ \frac{k^2}{2\mu} \right]^{1/3} |\xi_n|
\label{EAiry3D}
\end{equation}
and eigen-functions:
\begin{equation}
\rho_n (r) = \frac{1}{r} \mbox{Ai} \left( r \left[ 2 \mu k \right]^{1/3} - \xi_n \right).
\end{equation}
The ground state corresponds to the first zero of the Airy function, with energy $E_0 = 2.3381$,
 in $m_1 = m_2 = m = k = 1$ units.

\subsection{Meson variational wave function}
Previous studies have explored how to approach a variational wave function that can account for the meson energy \cite{carlson,stancu94,Val,Ca}. Here, we describe, in similar fashion, a family of 3-D variational wave functions that  approach this value and find analytical expressions for the expected variational parameter, energy and mean square radius. The general variational wave function takes the following form:
\begin{equation}
F_{\lambda, b} (r) =A e^{-\lambda r^b},
\end{equation}

\noindent where $A$ is a normalization  term, $b$ and $\lambda$ are variational parameters and $r$ is the relative coordinate of the two-body system.
We have found that using $b = 1.74716$ and $\lambda = 0.350132$, the wave function reproduces the exact minimum energy value within a 1\% error.  Based on this analysis, we fix the variational parameter $b = 3/2$ and leave the $\lambda$ parameter as the only free parameter. Then, the normalized meson variational wave function becomes:
\begin{equation}
F_\lambda (r) = \sqrt{\frac{3 \lambda^2}{2 \pi}} e^{-\lambda r^{3/2}}.
\label{FlambdaMeson}  
\end{equation}
Analytical results can be obtained for the properties of the meson for this form of the variational wave function. The value of $\lambda$ minimizing the energy is:
\begin{equation}
 \lambda_0=\frac{2\sqrt{k\mu}}{3},
\end{equation}
corresponding to an energy of 
\begin{equation}
E_0=\frac{3^{5/3} k \Gamma  \left( \frac{8}{3} \right)}{2^{7/3}(k\mu)^{1/3}},
\end{equation}
which is only around 3\% higher than the exact solution (see Table \ref{ELzero}).
The mean square radius for the meson can be obtained following the general definition:
\begin{equation}
\left \langle {r}_M^2  \right \rangle \equiv \left \langle \sum_{i=1}^2 \left( \vec{r}_i - \vec{R} \right)^2 .\right \rangle
\label{msr}
\end{equation}

\noindent  In the meson case with particles of mass $m_1$ and $m_2$, it becomes
\[
\langle r_M^2 \rangle_0 = \frac{m_1^2 + m_2^2}{(m_1 + m_2)^2}  \frac{3^{4/3}}{4} \Gamma \left( \frac{10}{3} \right),
\]
\noindent  where the subindex $_0$ is used everywhere to denote the isolated case.
We can extend this analysis for quarks of different masses, in particular considering the constituent masses of the $d,s,c$ and $b$ quarks, we set the ratio of their masses respect to the $u$ quark mass, taken as reference \cite{PDG}. In Table \ref{ELzero}, we show the corresponding optimal variational parameter ($\lambda_0$), the variational energy ($E_0$), the exact energy  ($E_{\mbox{exact}}$) and the variational radius ($\langle r_M^2 \rangle_0$)  for the different mass ratios. We denote $u$ mass as $m_1$, and $m_2$ corresponds to effective $d,s,c$ and $b$ masses.
These results will be used as reference when extending to a four-body system.
\begin{table}
\begin{center}
\begin{tabular}{| c | c | c | c | c |} \hline \hline

 $m_2/m_1$ & $\lambda_0$ & $E_0$ & $E_{\mbox{exact}}$ & $\left\langle r_M^2 \right\rangle_0$\\ \hline \hline

1 & 0.4714 & 2.3472 & 2.33811& 1.50255 \\ \hline

 1.44643 & 0.5125 & 2.2197 & 2.21106 & 1.38885\\ \hline

 4.6131 & 0.6043 & 1.9889 & 1.98118 & 1.52605\\ \hline

 14.0774 & 0.6441 & 1.9061 & 1.8987 & 1.73656\\ \hline

\end{tabular}
\caption{Optimal variational parameter and energy, the exact energy and mean square radius for  different mass ratios respect to the lightest $u$ mass, $m_1$.}
\label{ELzero}

\end{center}
\end{table}

\section{Four-body system}
Using the previous results, we can now extend our study to a four-body system composed of two identical $u$-like quarks ($qq$) and two identical antiquarks ($\bar Q \bar Q$), which  can be $d,s,c$ and $b$-like. To simplify the notation, we will refer to this four-body system by $qQ$.
At very low density, where the distance between quarks is large, the model must reproduce a system of two isolated mesons. In addition, as the density increases, the inter-particle separation becomes small allowing the interaction between the two mesons, which can be represented by quark exchange or a truly four-body interaction. All over, the Pauli blocking among identical particles must be enforced, allowing the system to behave like a free Fermi system at very high densities. Here, we rely on the String-Flip Model \cite{lenz,moniz,oka} which has been also used to study dense matter systems \cite{Piekarewicz,ayala}, to describe the dynamical transition among these regimes.

\subsection{String-flip Potential}
To model such a system, we need to capture the properties of the  strong interaction. In particular, 
the strong interaction of a many-body system can be well approached by considering that the quarks are connected by gluon flux tubes  \cite{isgur85} according to a configuration producing the lowest energy state of the whole system.\\ 
Let us denote the position of the particles by $\vec{r}_1$ and  $\vec{r}_2$ and the antiparticles by $\vec{r}_3$ and $\vec{r}_4$. The way to link the four particles of the system consistent with the QCD restrictions of color neutrality is two-fold:\\
 i) The system is composed of two mesons. Then, using the linear potential per pairs  $V \left( \vec{r}_i, \vec{r}_j \right) \equiv k \left\vert \vec{r_i} - \vec{r_j} \right\vert=k r_{ij}$, the total potential can be either
\begin{equation}
V_{m1}= V(\vec{r}_1,\vec{r}_3)+V(\vec{r}_2,\vec{r}_4)
\end{equation}
or
\begin{equation}
V_{m2}= V(\vec{r}_1,\vec{r}_4)+V(\vec{r}_2,\vec{r}_3).
\end{equation}

ii) The system is composed of a tetraquark. In this case, the shortest path linking the four particles is given by the Steiner-tree, which uses two auxiliary vectors (denoted by $\vec{k}$ and $\vec{l}$) to minimize the length of the configuration.  The vector $\vec{k}$ links the diquark sub-system and the vector $\vec{l}$ the anti-diquark one. The diquark and anti-diquark are considered to belong to the {\bf 3} and ${\bf \bar 3}$ representations of the color SU(3) symmetry, respectively. Thus, the total system form a color singlet state. The interaction among  particles in such representations is attractive with a strength a half of the quark-antiquark interaction. Since there is no direct link between particles but rather an interaction via auxiliary vectors, we use such strength as the same among all of them. This form of the potential has been shown to be consistent with the flux tube picture \cite{isgur85} and found, in the Lattice QCD framework, to follow a linear behavior at large distances \cite{potential}.
Thus, the potential becomes:
\begin{equation}
V_{4Q} = \sum_{i=1}^2 V ( \vec{k}, \vec{r}_i )/2+ \sum_{j=3}^4 V ( \vec{l}, \vec{r}_j )/2 + V(\vec{k},\vec{l})/2,
\label{PotTree}
\end{equation}
where the 1/2 factor accounts for the decreasing in the coupling strength.
When all these configurations are allowed, the potential of the system is chosen as the one producing the minimum potential energy in a given configuration:
\begin{equation}
V = min (V_{m1}, V_{m2},V_{4Q}).
\end{equation}

\subsection{Variational Wave function}
\noindent Generalizing the meson result  to include all the possible configurations at different densities, we propose the following variational wave function, which reproduces the two isolated mesons at very low densities, allows four-body correlations and incorporates the Fermi correlations as the density increases:
\begin{equation}
\Psi_{\lambda} = \Phi_{FG} \left( e^{-\lambda Q} \right),
\label{WFgeneral}
\end{equation}

\noindent where $\lambda$ is the single variational parameter, $Q$ is a function driven by the many-body potential, and $\Phi_{FG}$ is the free Fermi system wave function given by a product of Slater determinants, one for each color-flavor combination of quarks.
For a four-body system, the $Q$ function takes the following form, depending on the optimal potential at the given configuration, denoted by the sub-index:
\begin{equation}
Q_{m_1} =  r_{13} ^{3/2} +r_{24}^{3/2} ,
\end{equation}
\begin{equation}
Q_{m_2} =  r_{14}^{3/2} + r_{23}^{3/2} 
\end{equation}
or
\begin{equation}
Q_{4Q} =  r_{1k}^{3/2} + r_{2k}^{3/2} + r_{kl}^{3/2} + r_{3l}^{3/2} + r_{4l}^{3/2} ,
\end{equation}
where the 3/2 power is the resemblance of the exact solution of single pairs linked by a linear potential.

The Slater determinants are composed of single wave functions of a particle in a cubic box of side $L$ with Eigen-energy
\begin{equation}
E_n = \frac{\pi^2 \hbar^2}{2mL^2} \left( n_x^2 + n_y^2 + n_z^2 \right)
\label{Energiacaja}
\end{equation}

\noindent where $\quad n_x, n_y, n_z = 1,2,...$ We expect this form of the variational wave function to  be able to describe the departure from the two-mesons case into the four-body case, while including the identical particles correlations.

We can define a particle density parameter as a measure of the interparticle separation by:
$\rho = N/V= 4/L^3$, where $N$ is the number of particles of the system and $V$ is the box volume. For a fixed number of particles, the change in the particle density corresponds to modify the box size and correspondingly to the inter-particle separation.

\section{Energy determination}
The Hamiltonian of 4 particles of momentum $\vec{P}_i$ and mass $m_i$ interacting through the potential V is:
\begin{equation}
H = \sum_{i=1}^4 \frac{\vec{P}_i^2}{2m_i} + V. 
\end{equation}
Using the above variational wave function, the expectation value of each term in the  kinetic energy can be set as: 
\begin{equation}
 \frac{-1}{2 m_i Z(\lambda)} \int  \Psi_{\lambda} \partial_i^2 \Psi_{\lambda} dx
\end{equation}

\noindent where $Z(\lambda) \equiv \langle \Psi_{\lambda} \vert \Psi_{\lambda}  \rangle$.
Defining $\Psi_{\lambda} \equiv F_{\lambda} \Phi $, with $F_{\lambda} \equiv e^{-\lambda Q} $, then
\begin{equation}
\partial_i^2 \Psi_{\lambda}= F_{\lambda} \partial_i^2 \Phi_{FG}+ 
   \partial_i^2 F_{\lambda}  \Phi_{FG} + 2  \partial_i F_{\lambda}  \partial_i \Phi_{FG}. 
\end{equation}
The first term leads to the noninteracting particle kinetic contribution ($T_{FG}$) and the remaining terms, after integrating by parts, to:
\begin{equation}
 \frac{1}{2m_i Z(\lambda)} \int  \Psi_{\lambda}^2 [\partial_i ln F_{\lambda}]^2 dx.
\end{equation}
Thus, using the exponential structure of $F_{\lambda}$, this term corresponds to the expectation value of the following function: 
\begin{equation} 
W_i \equiv\frac{\lambda^2}{2m_i} \left[ \partial_i  Q \right]^2,  \ \  W\equiv \sum_i^4 W_i,
\label{wkin}
\end{equation}
where the sum runs over the coordinates of the four particles. This term can be seen as the correction to the kinetic energy due to the interaction. Thus, the expectation value of the Hamiltonian becomes \cite{Piekarewicz}:
\begin{equation} 
\left\langle H \right\rangle_{\lambda} = T_{FG} + \left\langle W \right\rangle_{\lambda} + \left\langle V \right\rangle_{\lambda}.
\end{equation}

\noindent This structure is useful to find the variational parameter  $\lambda$ that minimizes it:
\begin{equation} 
 \frac{\partial \left\langle H \right\rangle_{\lambda}}{\partial \lambda} = 0.
\end{equation}
The free kinetic part is given by the sum of the energies of the particles in the box as given by Eqn. (\ref{Energiacaja}) and is independent of $\lambda$.
In order to evaluate $<W>$, we need to know the particles coordinates dependence of $Q$. For the meson configuration, $Q$ is explicitly defined in terms of them, while for the tetraquark it involves the vectors $\vec{k}$ and $\vec{l}$, whose dependence on the coordinates are not analitically known, in this case numerical determination is used (see Appendix). 

\section{Results}
The expectation values of $W$ and $V$ involves the integration over 12 variables and can be computed using Monte Carlo techniques which, by relying in the importance sampling,  turns the integrals into an evaluation of the average of the values of the observable, in particular we use a Metropolis algorithm for the sampling, driven by the square of the wave function.
 We computed the properties of three systems. The results we present are organized as follows:

 \begin{itemize}
 \item \emph{Two mesons}. In this case, only meson configurations are allowed in the potential. We characterize the meson properties, comparing the results with the expected from the analytical solution in the isolated limit and the departure from them as density increases. 
 \item \emph{Tetraquark }.  In this case, only tetraquark configurations are allowed in the potential. We characterize the tetraquark properties in the isolated limit and the departure from them as density increases.
 \item \emph{Mixed}.  In this case, all the possible configurations are allowed in the potential. We characterize the meson and tetraquark properties and their modifications due to the presence of the other configuration.
\end{itemize}
For all the above cases we explored the mass effect by considering the two quarks to be light ($u$)and the antiquarks having variable mass to resemble the  $m_{d,s,c,b}/m_u$ ratios.
Results are presented for several light-quarks to heavy-antiquarks mass ratios whenever  they are found to be relevant.

\subsection{Variational parameter and energy}  
In order to know the best approach of the variational wave function to the exact solution, we find the optimal value of the variational parameter that minimizes the total energy at a given density.  We repeat the procedure for a set of values of the density to know the corresponding evolution of the variational parameter.

\begin{figure}
\centerline{\epsfig{file=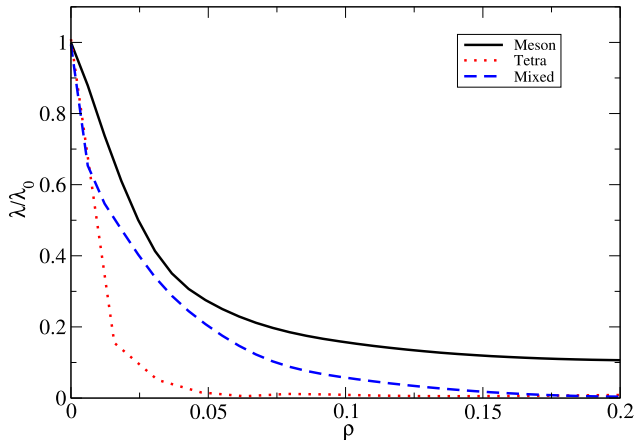,angle=0,width=9.5cm}}
\vspace{-0.1in} \caption{ Variational parameter of the two mesons, tetraquark and mixed systems as a function of the density, for a $ud\equiv uu\bar d \bar d$ mass assignment. They are normalized to the exact solution in the isolated  meson case $\lambda_0$.}
\label{fig1m}
\end{figure}

In Figure  \ref{fig1m}, we show the optimal values of the variational parameter 
of the two mesons, tetraquark and mixed systems as a function of the density, for a $ud\equiv uu\bar d \bar d$ mass assignment. The behavior for a light-heavy system ($ub$) is very similar (not shown) thus, for the sake of clarity, we only exhibit this case.
The results are normalized to the exact solution in the isolated  meson case $\lambda_0$. We observe that they exhibit a smooth evolution from their asymtotic value at low density decreasing to nearly zero at high energy densities, signing the diminishing effect of the interaction. 

\begin{figure}
\centerline{\epsfig{file=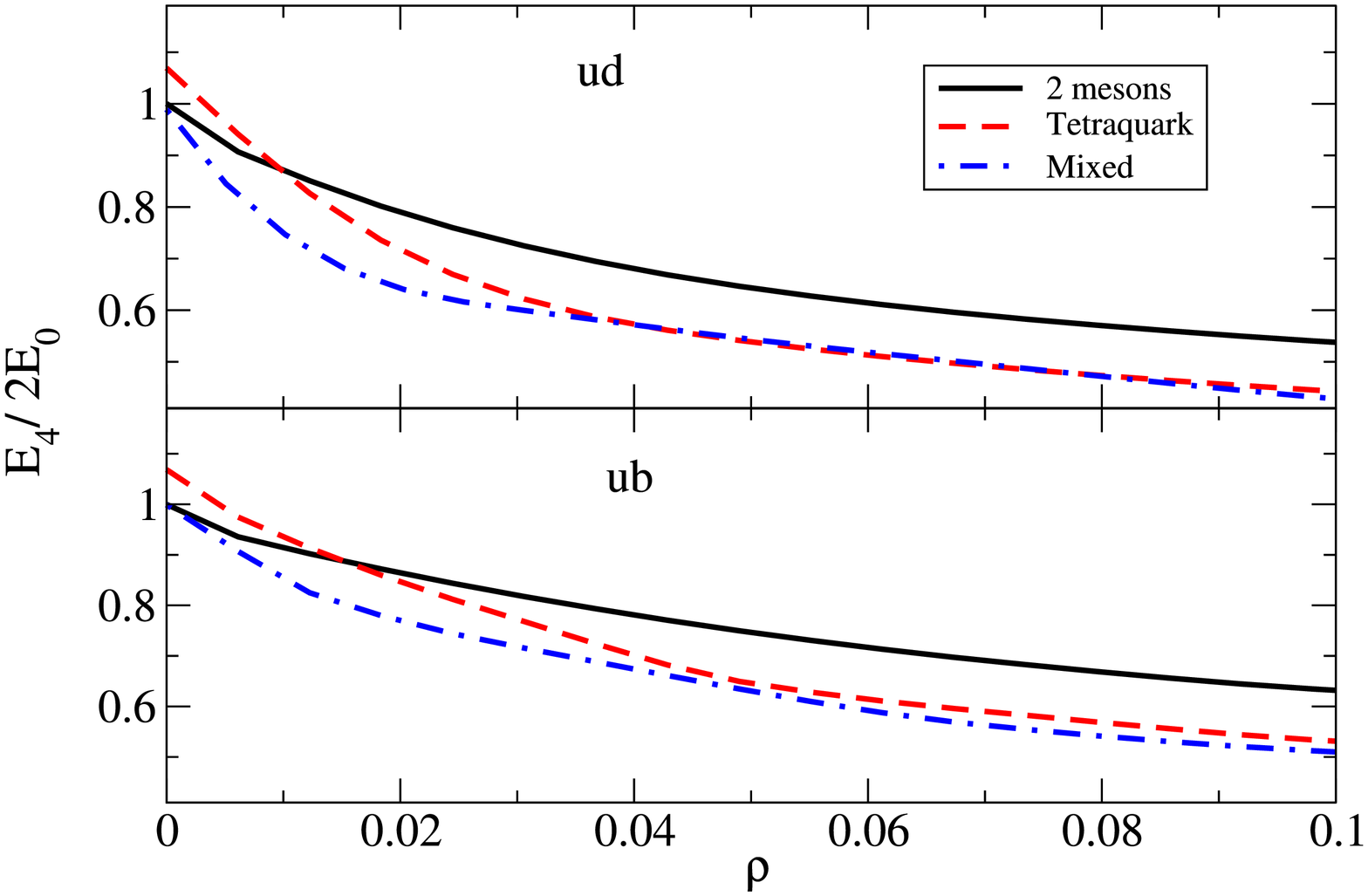,angle=0,width=9.5cm}}
\vspace{-0.1in} \caption{ Energy of the two mesons, tetraquark and mixed systems as a function of the density, normalized to the exact solution of two isolated mesons, 2$E_0$. The upper (lower) panel corresponds to a $ud$ ($ub$) system.}
\label{energy}
\end{figure}

In Figure \ref{energy}, we have plotted  the energy of the system ($E_4$), normalized to the exact solution in the isolated meson case ($2E_0$) for the three cases: two mesons, tetraquark and mixed systems, as a function of the particle density. The upper (lower) panel corresponds to a $ud$ ($ub$) system. We observe that at low density we reproduce the exact result for the two mesons system and the tetraquark system energy is higher. As the density increases, this breach closes and eventually is inverted, signing that the tetraquark state is more likely at high density. The mixed system turns out to be the one with the lowest energy as it takes advantage of both possible configurations.

\subsection{Radial distribution}
\begin{figure}
\centering{\epsfig{file=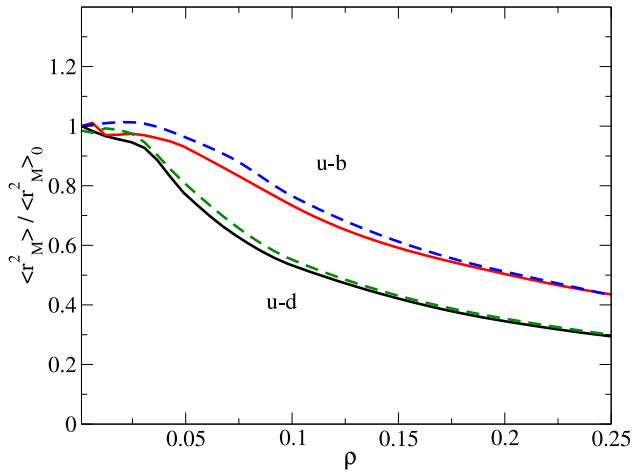,angle=0,width=10.cm}}
\vspace{-0.1in} \caption{Meson mean square radius as a function of density, normalized to the isolated case, for light-light (ud) and light-heavy (ub) mass ratios. The solid and dashed line correspond to the meson and mixed system respectively.}
\label{fig3m}
\end{figure}

\begin{figure}
\centering{\epsfig{file=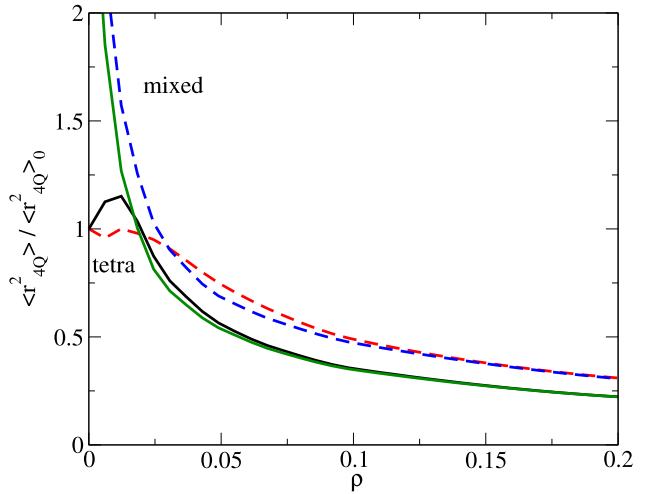,angle=0,width=10.5cm}}
\vspace{-0.1in} \caption{Tetraquark mean square radius as a function of density, normalized to the isolated case. The solid (dashed) line corresponds to a $ud$ ($ub$) mass ratios, for the Tetraquark and mixed systems.}
\label{fig3q}
\end{figure}

Once the optimal variational wave function and energy were charaterized, we proceed to compute a set of observables. In particular, the mean square radius (MSR) evolution was obtained, using the optimal pairing information.
The mean square radius for the meson or the tetraquark can be obtained following the general definition, Eqn. (\ref{msr}), where the sum runs over the corresponding number of particles.
In Figures  \ref{fig3m} and \ref{fig3q}, we show the density evolution of the MSR for mesons and tetraquarks respectively, normalized to the corresponding value in the isolated case.

The meson MSR evolution is found to be mass ratio dependent. In fact, the light-light combination  ($ud$) decreases faster than the light-heavy case ($ub$), while the dependence on being in presence of possible tetraquark configurations is mild (solid vs. dashed line) in both cases.

The tetraquark MSR evolution exhibits a mass ratio dependence. The light-light system (solid line) increases in the low density regime and then rapidly decreases. The light-heavy system (dashed line) remains almost unchanged  for a relatively large density region and slowly decreases. For both cases we show the modification when considering the presence of the mesons (mixed). Both become large at low density and rapidly decrease as the density increases, approaching the pure tetraquark result. 
It is very interesting to find out that during the increase of the density, the radial distribution suffers not only contractions but also expansions, which are a result of a competition between the reduction of the interparticle separation and the decrease of the confining potential effect, as it is multiplied by the density decreasing variational parameter.

In Table \ref{4Qzero}, we  show the tetraquark energy ($E_{{4Q}_0}$) and MSR  ($\left\langle r_{4Q}^2 \right\rangle_0$)  in the limit of zero density, for the different mass ratios. Twice the meson energy  ($2E_0$) is also exhibit as reference. 

\begin{table}
\begin{center}
\begin{tabular}{| c | c | c | c |} \hline \hline

 $m_2/m_1$ & $2E_0$ & $E_{{4Q}_0}$ & $\left\langle r_{4Q}^2 \right\rangle_0$\\ \hline \hline

1 & 4.6944            & 5.02   $\pm$    0.02  & 13.9    $\pm$    0.1  \\ \hline

 1.44643 & 4.4394 &  4.77     $\pm$  0.02 & 13.2     $\pm$   0.2\\ \hline

 4.6131  & 3.9778 & 4.24    $\pm$   0.02  &12.5  $\pm$    0.3 \\ \hline

 14.0774 & 3.8122 & 4.07    $\pm$    0.02 &  13.1   $\pm$    0.2  \\ \hline

\end{tabular}
\caption{Two mesons energy, Tetraquark energy  and mean square radius for  different mass ratios respect to the lightest $u$ mass, $m_1$ in the zero density limit.}
\label{4Qzero}
\end{center}
\end{table}

\subsection{Correlation functions}
Another very useful observable to characterize the properties of the system is the two-particle correlation function  \cite{walecka}. It measures the probability of finding two particles at a relative distance $r$ from each other:
\begin{equation}
g(r) \equiv \frac{V}{4 \pi r^2 N^2}  \left\langle \sum_{i < j = 1}^N \delta\left( \vec{r} - \vec{r}_{ij} \right) \right\rangle
\end{equation}

\begin{figure}
\centerline{\epsfig{file=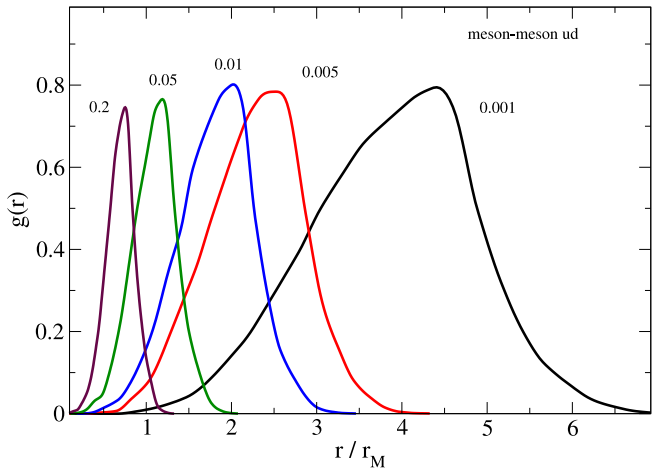,angle=0,width=10.cm}}
\vspace{-0.1in} \caption{Meson-meson correlation function at several densities (inner labels). Corresponding to a light-light ($ud$) system, for the two mesons case.}
\label{mmud}
\end{figure}

\begin{figure}
\centerline{\epsfig{file=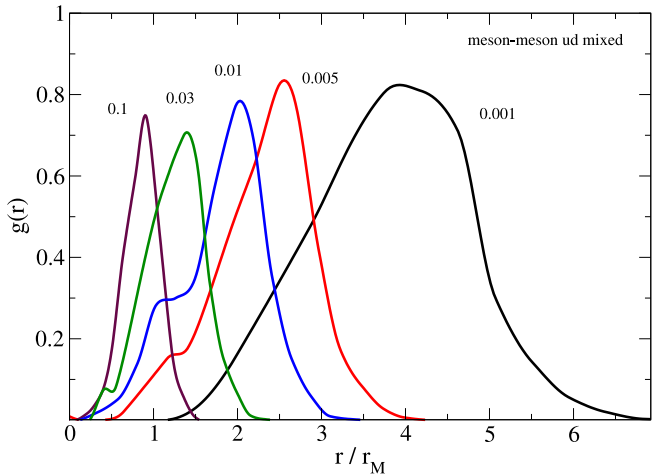,angle=0,width=10.cm}}
\vspace{-0.1in} \caption{Meson-meson correlation function at several densities (inner labels). Corresponding to a light-light ($ud$) system, for the mixed case.}
\label{mmmixud}
\end{figure}

In Figures \ref{mmud} and \ref{mmmixud} we show the meson-meson correlation function at several densities for the two-mesons and mixed case, respectively. We exhibit the results corresponding to a light-light ($ud$) system but a similar behavior is observed in the other mass ratio combinations. The comparison between both Figures shows that there is a modification in the meson-meson correlation function by the presence of the tetraquark state at intermediate densities, a bump develops in the near tail of the correlation function, driven by the diquark formation (see below), and fades out as the density increases and all the quark correlations vanish. This may be an alternative observable to have indirect evidence of the tetraquark formation. Note that at low density the separation between the two mesons approaches the result obtained by Lenz et al \cite{lenz} of 4.66 $r/r_M$.\\
For the two meson case, the quark-quark correlation function at low density is similar to the meson-meson correlation function (not plotted) as the  relevant separation distance is driven by the separation between mesons, each one containing a quark.

\begin{figure}
\centerline{\epsfig{file=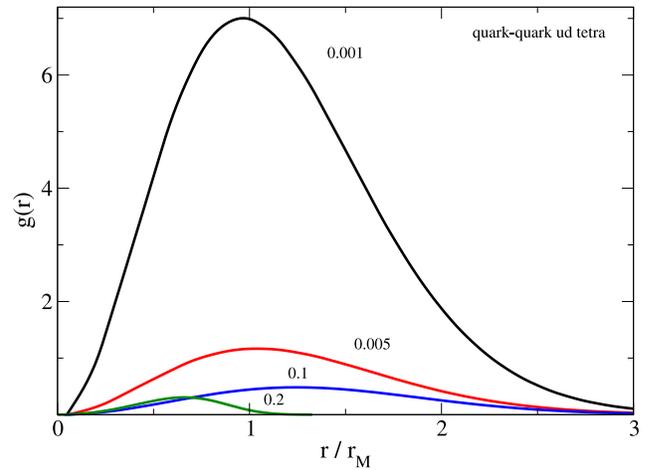,angle=0,width=10cm}}
\vspace{-0.1in} \caption{Quark-quark correlation function at several densities (inner labels), for a light-light ($ud$) system in the tetraquark case.}
\label{qqtetraud}
\end{figure}

In Figure \ref{qqtetraud}, we show the quark-quark correlation function at several densities (inner labels) for a light-light ($ud$) system, when only tetraquark configurations are allowed.
The diquark formation at low densities is exhibited and, as the density increases, the correlation drops signing that, although the tetraquark configuration can be identified, the quarks are no longer correlated. Note that the diquark size is similar to the bump observed in the meson-meson correlation function (Figure \ref{mmmixud}) when tetraquarks are also formed. At low densities the fraction of tetraquarks formed respect to the two mesons is small and has no major effect, it is only in the intermediate density region where this feature becomes relevant.

\begin{figure}
\centerline{\epsfig{file=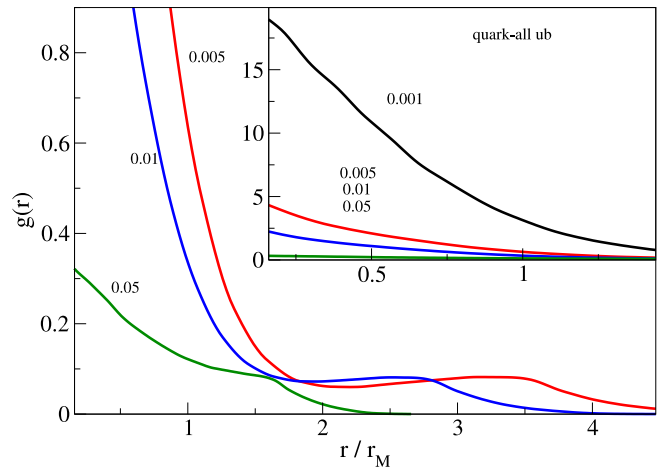,angle=0,width=10cm}}
\vspace{-0.1in} \caption{Correlation between a quark and all the particles regardless of the quantum numbers. Two mesons system.}
\label{qallub}
\end{figure}

\begin{figure}
\centerline{\epsfig{file=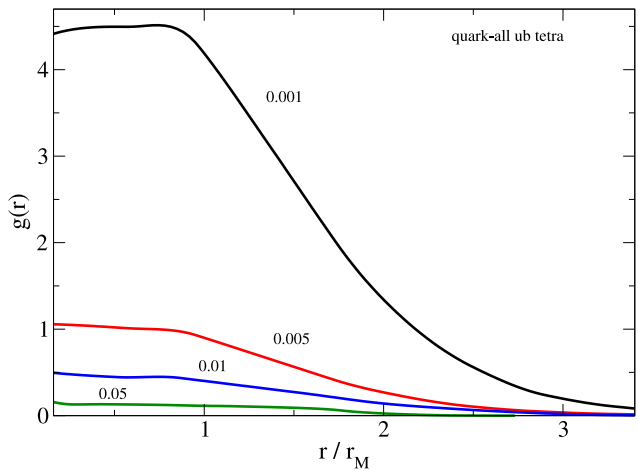,angle=0,width=10cm}}
\vspace{-0.1in} \caption{Correlation between a quark and all the particles regardless of the quantum numbers. Tetraquark system.}
\label{qallubtetra}
\end{figure}

In Figures \ref{qallub} and \ref{qallubtetra}, we show the correlation between a quark and all the particles regardless of the quantum numbers, for two mesons and tetraquark systems respectively. Here, we exhibit the case of the light-heavy ($ub$) combination. Although the general features are similar for other combinations the specific parameters are not. This will be exemplified below, when discussing the zero density limit.

In the zero density limit, the normalized tetraquark system correlation function for the diquark (quar-quark) and tetraquark (quark-all)  can be parameterized by:
\begin{equation}
g(r)_{q-q}= A_0 r^2 e^{-r^{A_2}/A_1^2}
\end{equation}
and
\begin{equation}
g(r)_{q-all}= A_0 (1+ r/A_1) e^{-r^2/A_2}
\end{equation}

In Table \ref{diquarkpar}, we show the value of the parameters of the diquark and quark-all correlation functions  for the light-light and light-heavy cases. Note that the dependence on $r$  of  the diquark correlation function is close to  others suggested in the literature \cite{formfactor,bicudo13} but in our case it is dynamically obtained. The short distance region shows the competition between the Pauli blocking among the identical particles and the attractive potential.
\begin{table}
\begin{center}
\begin{tabular}{| c | c | c |c |  } \hline \hline
System & $A_0$ & $A_1$ &$A_2$\\ 
\hline
Quark-quark&&&\\
\hline
 ud &
 0.64&
 1.24&
 1.51\\
 ub &  
 1.13&
 1.1&
1.47    \\
 \hline
Quark-all&&&\\
 \hline
 ud & 
  0.24 &
 1.00 &
 4.60\\
ub&
 0.31 &
 0.98&
 3.19 \\
\hline
\end{tabular}
\caption{Diquark and quark-all  correlation functions parameters for $ud$ and $ub$-like mass ratios.}
\label{diquarkpar}
\end{center}
\end{table}

\begin{figure}
\centerline{\epsfig{file=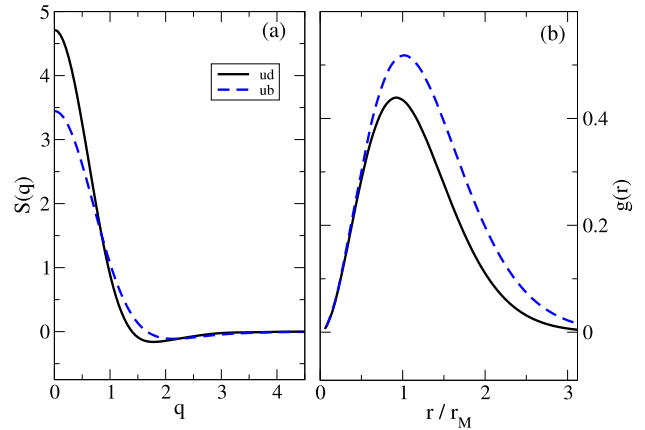,angle=0,width=9.cm}}
\vspace{-0.1in} \caption{Diquark static structure factor $S(q)$  (a) and correlation function (b) in the zero density limit, for $ud$ (solid line) and $ub$ (dashed line) tetraquark systems.}
\label{sqdiquark}
\end{figure}

The static structure factor $S({\bf q})$ can be obtained as the Fourier transform of the correlation function $g({\bf r})$. That is
\begin{equation}
S({\bf q})= 1+ \frac{N}{V}\int d^3r \ g({\bf r}) \ e^{-i{\bf q}\cdot {\bf r}}.
\end{equation} 
In our case we consider the correlation to depend on the magnitude of ${\bf r}$. 
In Figure \ref{sqdiquark}, we display the diquark static structure factor $S(q)$  (a) and correlation function (b)  in the limit of zero density for both $ud$ (solid line) and $ub$ (dashed line) tetraquark systems.

\subsection{Dynamical recombination}
The  four quark recombination can have important effects in systems where they are produced very close in space. An example can be found in the $WW \to qq\bar Q \bar Q$ decay, whose spatial separation at LEP2 energies is around 0.1 fm \cite{wmass}. Although in the perturbative regime the recombination is small, in the non-perturbative regime the effect may be important.
The typical scenarios to estimate the recombination are: Considering spherical or elongated bags color sources and the reconnection is proportional to the overlap of two color sources;
strings considered as vortex lines where reconnection takes place when the core of the two string pieces cross each other \cite{ww}. The recombination between the two mesons configurations is usually implemented \cite{wmass,ww}.
In our case, we would have, in addition, recombinations similar to tetraquark states which eventually freeze out to two mesons.
 Here, we characterize the modification to the two mesons recombination induced by the presence of the tetraquark state.\\
 The model requires that the formed clusters must be colorless. The flipping from one configuration to another, driven by the minimal potential energy restriction, is a measure of the dynamical recombination property of the system. We define the recombination probability for the two mesons case  by:  
\begin{equation}
Pr_{2m}= \frac{N(V_{m1} \leftrightarrow V_{m2}) }{N(V_{m1}) + N(V_{m2})}
\end{equation}
where $N(V_i \to V_j)$ denotes the number of flippings from the $i$ to the $j$ configuration, and $N(V_i)$ is the number of times the system visits the $i$ configuration. If we consider the tetraquark configuration as another possible configuration contributing to the recombination, the probability becomes:
\begin{eqnarray}
Pr_{mix}=& \left[
N(V_{m1} \leftrightarrow V_{m2}) +N(V_{m1} \leftrightarrow V_{4Q})+ \right. \nonumber \\
&\left.N(V_{m2} \leftrightarrow V_{4Q})+ N(V_{4Q} \to V_{4Q})
\right]/ \nonumber \\
&\left[N(V_{m1})+N(V_{m2}) + N(V_{4Q}) \right]
\end{eqnarray}

\begin{figure}
\centerline{\epsfig{file=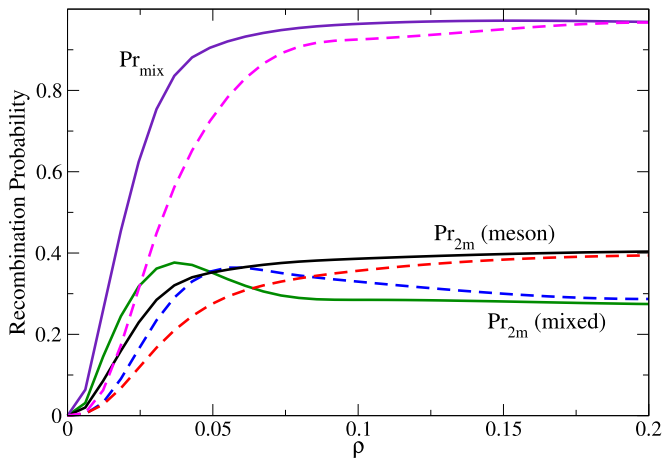,angle=0,width=10cm}}
\vspace{-0.1in} \caption{Recombination probability as a function of the density. The probability for only two mesons recombination ($Pr_{2m}$)  is exhibited for the two mesons and mixed cases (see inner labels). The probability when tetraquarks are accounted  as a recombination stage ($Pr_{mix}$) corresponds to the mixed case. The solid  and dashed lines are for $ud$  and $ub$ systems respectively.}
\label{recombination}
\end{figure}

In Figure \ref{recombination}, we show the probabilities as a function of the density. 
The probability for only two mesons recombination ($Pr_{2m}$)  is exhibited for the two mesons and mixed cases (see inner labels). The probability when tetraquarks are accounted  as a recombination stage ($Pr_{mix}$) corresponds to the mixed case. The solid  and dashed lines are for $ud$  and $ub$ systems respectively.

\begin{figure}
\centerline{\epsfig{file=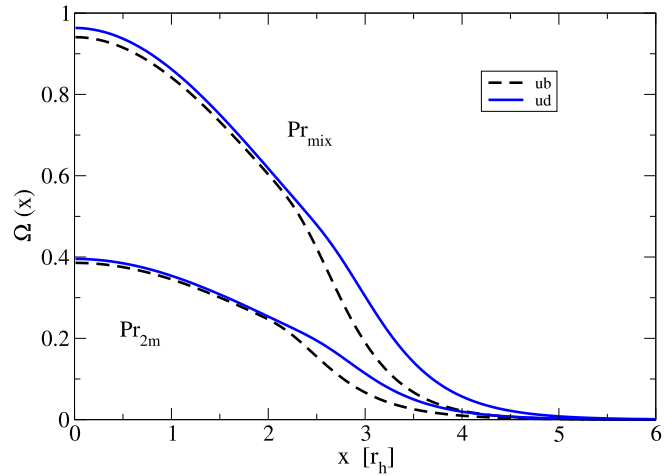,angle=0,width=10cm}}
\vspace{-0.1in} \caption{Color strength due to the recombination probability  in the two mesons and mixed cases as a function of the radial size of the system (in units of the meson radius). The solid  and dashed lines are for $ud$  and $ub$ systems respectively.}
\label{omega}
\end{figure}

A qualitative estimate of this effect can be exhibited considering a simple expansion model, such that we can evolve the system along the density profile. We can define the color strength function by:
\begin{equation}
\Omega (x)\equiv P_{frag}(t) Pr(x)
\end{equation} 
where $P_{frag}(t)=exp(-t^2/\tau^2_{frag})$ is the  probability that the system has not yet fragmented, with $\tau_{frag}\approx 3r_h$  the proper lifetime, taken as three times the meson radius $r_h$ \cite{ww}.
The recombination probability $Pr(\rho)$  can be set in terms of  the radius of the sphere ($x$) corresponding to the given density $\rho=3N/4\pi r_h^3 x^3$. In Figure \ref{omega}, we plot the evolution of the color strength as a function of the radial size of the system, in units of the meson radius. The case when considering the tetraquark and meson recombination ($Pr_{mix}$) largely modifies the estimate when the tetraquark state is not included ($Pr_{2m}$). The flavor composition modifies the  behavior at intermediate distances, and the total effect becomes negligible after four times the hadron radius,  corresponding to nearly the size of the end of the overlap of two hadrons.

\subsection{Tetraquark potential}
\begin{figure}
\centerline{\epsfig{file=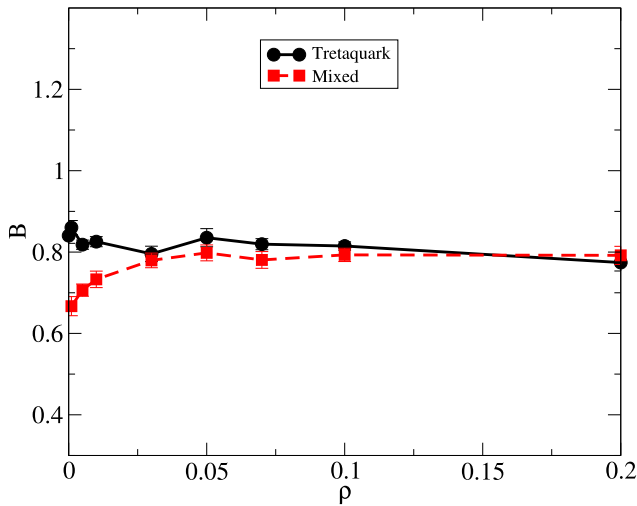,angle=0,width=11cm}}
\vspace{-0.1in} \caption{Slope ($B$) of the effective tetraquark linear potential as a function of density, for the pure tetraquar system (solid line) and meson- tetraquark mixed system (dashed line).}
\label{linearpotential}
\end{figure}

The tetraquark potential depends on the quarks positions and two auxiliary vectors, placed in such away that the total length linking the quarks is the shortest one.  These auxiliary vectors are modified in a non-trivial way whenever a single quark changes its position. An effective
behavior of the potential can be set as linear respect to the invariant length, $R\equiv \sqrt{\sum r^2_{ij}}$:
\begin{equation}
V(R)=R_0+BR
\end{equation}
where $R_0$ is the value of the potential at zero distance, which is expected to be modified by the short distance coulomb-like correction \cite{potential}. In Figure \ref{linearpotential}, we show the slope $B$ of the linear behavior from the simulation as a function of density for the case when only tetraquarks are allowed to form (solid line) and when mesons and tetraquarks are allowed to form (dashed line). The slope, in the tetraquark case  shows small dependence on the density, and $B(\rho \approx 0) =0.84\pm 0.02$. In the mixed case there is a significant density dependence, starting below  the corresponding value at zero density ($B(\rho \approx 0)=0.67\pm 0.02$) and as the density increases they approach to each other. This behavior is similar when considering different flavor  systems.
 
\begin{figure}
\centerline{\epsfig{file=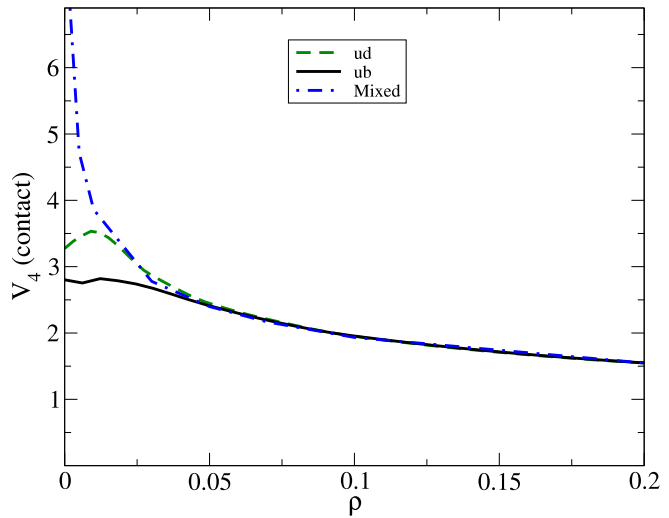,angle=0,width=11cm}}
\vspace{-0.1in} \caption{Effective tetraquark contact potential as a function of density, for the pure tetraquar system (solid and dashed lines for $ub$ and $ud$ flavors respectively) and meson- tetraquark mixed system (dotted line).}
\label{contact}
\end{figure}

In Figure \ref{contact}, we show the effective four-body contact potential, $V_4(contact)$,  as a function of density, determined as the average potential irrespective of the length linking the particles.  The pure tetraquar system (solid and dashed lines for $ub$ and $ud$ flavors respectively) exhibit a dependence on the mass ratio in the low energy regime and, as density increases, they become similar. The meson-tetraquark mixed system (dotted line) has no dependence on the mass ratio (labeled as mixed), but is is very sensitive to the presence of the mesons, becoming very large in the low energy regime. This behavior corresponds to the increasing distance among quarks which are influenced by the two mesons configuration that is trying to bring them apart.

\section{Conclusions}

We have performed a Monte Carlo simulation of two identical light quarks $qq$ and two identical antiquarks $\bar Q\bar Q$, considering three possible structures: two mesons, tetraquark  and mixed configurations. We determined whether it is energetically more favorable to form a tetraquark or two mesons and the mixing among them, as a function of the particle density.
Several light-quarks to antiquarks mass ratios were used and exhibited whenever  they were found to be relevant.
The meson MSR evolution was found to be mass ratio dependent. In fact, the light-light combination  ($ud$) decreases faster than the light-heavy case ($ub$), while the dependence on being in presence of possible tetraquark configurations is mild. 
The tetraquark MSR evolution also exhibits a mass ratio dependence, and turns out to be very sensitive to the presence of the mesons.

We have shown that there is a modification in the meson-meson correlation function by the presence of the tetraquark state at intermediate densities, where a bump develops in the near tail of the correlation function, driven by the diquark formation.
To gain insight into the features of the tetraquark state, we computed the quark-quark correlation function at several densities. We observed the diquark formation at low densities and the drop on the correlation as the density increased. The quantum correlations among identical particles appeared as Pauli blocking in the short distance regime.
In the zero density limit, we determined the correlation functions between quarks and the correlation between a quark and  the rest of the particles. A parameterization was found, which is useful to compute additional  static properties, in particular we computed the diquark static structure factor, which exhibited a small mass ratio dependence.

We did track the dynamical flipping among configurations and determined the recombination probability evolution as a function of the particle density. We have shown that the probability is largely affected when considering the tetraquark as an intermediate recombination state. 
Using a simple model evolution of the system along the density profile we have shown that the recombination effects remain important beyond a half the overlap of two mesons.

The  linear behavior of the four-body potential on the invariant length linking the quarks was analyzed and found  that the presence of a mixed state is reflected in the decreasing strength of the slope. We determined an effective four-body contact potential as the average strength, irrespective of the length linking the particles. This is very sensitive to the presence of the mesons and when they are not considered  there is a dependence on the mass ratio in the low energy regime.
 
The approach we have followed to the tetraquark system is intended to exhibit the gross features. Certainly, the short distance information and more elaborated interactions should have an effect on the system energy and the corresponding spectrum. In this work, we have limited ourselves to the case of identical quarks and identical antiquarks. Different particles effects should be reflected in the high density regime due to the absence of Fermi correlations.

\appendix*
 \section{Kinetic term}
The kinetic term $W$ Eqn. (\ref{wkin}) is evaluated as follows: 

{\it Meson}: In this case we have two configurations, corresponding to $Q_{m_1}$ and $Q_{m_2}$. Let us illustrate the form of the contribution from the coordinate $x$ of  particle 1, by taking the derivative of the first element of $Q_{m_1}$:
\[
\partial_{x_1} r_{13}^{3/2} = \frac{3}{2} \frac{ ( x_1 - x_3 )}{r_{13}^{1/2}}.
\]
Thus, if we compute the contribution for the 3 coordinates and add their squares, we get:
\[
\left[ \partial_{x_1} Q_{m_1} \right]^2+\left[ \partial_{y_1} Q_{m_1} \right]^2+\left[ \partial_{z_1} Q_{m_1} \right]^2 = \frac{9}{4} r_{13}
\]
Analogous results are obtained for the other particles, adding all them up we get:
\begin{equation}
\left\langle W (Q_{m1})\right\rangle_\lambda  = \frac{9 \lambda^2}{8} \left( \frac{ <r_{13}>}{\mu_{13}} +\frac{ <r_{24}>}{\mu_{24}}  \right) 
\end{equation}
where $\mu_{ij}$ is the corresponding reduced mass.
The $Q_{m_2}$ configuration kinetic term is obtained by the $3\leftrightarrow 4$ exchange. 

{\it Tetraquark}: In this case we have  two additional vectors $\vec{k}$ and $\vec{l}$, whose explicit dependence on the coordinates of the quarks and antiquarks are unknown.
Let us illustrate this implication in the contribution from the coordinate $x$ of  particle 1:
\begin{eqnarray}
& \partial_{x_1}  Q_{4Q} = \partial_{x_1} \left( r_{1k}^{3/2} + r_{2k}^{3/2} + r_{kl}^{3/2} + r_{3l}^{3/2} + r_{4l}^{3/2} \right) \nonumber\\
&= \frac{3}{2} r_{1k}^{1/2} \partial_{x_1} r_{1k} +...
\end{eqnarray}
The first term derivative can be set as:
\begin{eqnarray}
\partial_{x_1} r_{1k}& =& \left[ \left(x_1 - x_k \right) \left( 1 - \partial_{x_1} x_k \right) - \left(y_1 - y_k \right)  \partial_{x_1} y_k \right.\nonumber\\
&&
\left. -(z_1 - z_k ) \partial_{x_1} z_k \right]/
r_{1k}
\end{eqnarray}
where we do not know the derivative of the $\vec{k}$ coordinates respect to the $x_1$ coordinate. The derivatives are numerically evaluated as $
\partial_{x_1} x_k \sim \Delta x_k/\Delta x_1 $. Analogous terms are evaluated for all the rest of the particles coordinates.

\begin {acknowledgments}
We acknowledge the support of DGAPA-PAPIIT UNAM, grant IN106913. We thank  Dr. J. Piekarewicz and Dr. A. Ayala for very useful observations.
\end {acknowledgments}

\end{document}